\documentclass[5pt, twocolumn]{aastex62}

\received{February 2, 2019}
\revised{March 2, 2019}
\accepted{\today}
\submitjournal{ApJ}

\shorttitle{Coma Ber Tail}
\shortauthors{Tang et al.}

\begin{document}

\title{Discovery of Tidal Tails in Disrupting Open Clusters: Coma Berenices
        and a Neighbor Stellar Group}
	
\author{Shih-Yun Tang} 
\affiliation{Department of Physics, National Central University, 
        300 Zhongda Road, Zhongli, Taoyuan 32001, Taiwan}

\author{Xiaoying Pang$^*$}
\affiliation{Xi'an Jiaotong-Liverpool University, 111 Ren'ai Road, Dushu Lake Science 
        and Education Innovation District, Suzhou 215123, Jiangsu Province, P. R. China. Xiaoying.Pang@xjtlu.edu.cn}
\affiliation{Shanghai Institute of Technology, 100 Haiquan Road,
        Fengxian district, Shanghai 201418, P.R. China.}
\affiliation{Shanghai Key Laboratory for Astrophysics, Shanghai Normal University, 
        100 Guilin Road, Shanghai 200234, P.R. China}

\author{Zhen Yuan}
\affiliation{Key Laboratory for Research in Galaxies and Cosmology, 
        Shanghai Astronomical Observatory, Chinese Academy of Sciences,
        80 Nandan Road, Shanghai 200030, P.R. China}

\author{W.~P. Chen}
\affiliation{Graduate Institute of Astronomy, National Central University, 
        300 Zhongda Road, Zhongli, Taoyuan 32001, Taiwan}
\affiliation{Department of Physics, National Central University, 
        300 Zhongda Road, Zhongli, Taoyuan 32001, Taiwan}
        
\author{Jongsuk Hong}       
\affiliation{Kavli Institute for Astronomy and Astrophysics, Peking University,
        Yi He Yuan Lu 5, HaiDian District, Beijing 100871, P.R. China}
        
\author{Bertrand Goldman}
\affiliation{Max Planck Institute for Astronomy, Konigstuhl 17, D-69117 Heidelberg,
        Germany}
\affiliation{Observatoire astronomique de Strasbourg, Universit\'e de Strasbourg -
        CNRS UMR 7550, 11 rue de l'Universit\'e, 67000, Strasbourg, France}

\author{Andreas Just}
\affiliation{Zentrum f\"ur Astronomie der Universit\"at          
        Heidelberg, Astronomisches Rechen-Institut,                  
        M\"onchhofstr. 12-14, 69120 Heidelberg, Germany}
        
\author{Bekdaulet Shukirgaliyev}
\affiliation{Zentrum f\"ur Astronomie der Universit\"at 
        Heidelberg, Astronomisches Rechen-Institut, 
        M\"onchhofstr. 12-14, 69120 Heidelberg, Germany}
\affiliation{Fesenkov Astrophysical Institute, Observatory str. 23, 050020 Almaty, Kazakhstan}
\affiliation{Faculty of Physics and Technology, Al-Farabi Kazakh National University, Al-Farabi ave.
        71, 050040 Almaty, Kazakhstan}

\author{Chien-Cheng Lin}
\affiliation{Max Planck Institute for Astronomy, Konigstuhl 17, D-69117 Heidelberg,
        Germany}
\affiliation{Institute for Astronomy, University of Hawaii, 2680 Woodlawn Drive, Honolulu, HI 96822, USA}

\begin{abstract}
We report the discovery of tidal structures around the intermediate-aged ($\sim$ 700--800~Myr), nearby
($\sim85$~pc) star cluster Coma Berenices. The spatial and kinematic grouping of stars is determined 
with the {\it Gaia} DR2 parallax and proper motion data, by a clustering analysis tool, \textsc{StarGO},
to map 5D parameters ($X, Y, Z$, $\mu_\alpha \cos\delta, \mu_\delta$) onto a 2D neural network. 
A leading and a trailing tails, each with an extension of $\sim50$~pc are revealed for the first 
time around this disrupting star cluster. The cluster members, totaling to $\sim115^{+5}_{-3}\,\rm{M_\sun}$,
are clearly mass segregated, and exhibit a flat mass function with $\alpha \sim 0.79\pm0.16$, 
in the sense of $dN/dm \propto m^{-\alpha}$, where $N$ is the number of member stars and $m$ is stellar 
mass, in the mass range of $m=0.25$--$2.51~{\rm M_\sun}$. Within the tidal radius of $\sim$6.9~pc, 
there are 77 member candidates with an average position, i.e., as the cluster center, of 
R.A.= 186.8110~deg, 
and decl.= 25.8112~deg, and an average distance of 85.8~pc. Additional 120 member candidates reside 
in the tidal structures, i.e., outnumbering those in the cluster core. The expansion of escaping members 
lead to an anisotropy in the velocity field of the tidal tails. Our analysis also serendipitously 
uncovers an adjacent stellar group, part of which has been cataloged in the literature. We identify 
218 member candidates, 10 times more than previously known. This star group is some 65~pc away from, 
and $\sim400$~Myr younger than, Coma Ber, but is already at the final stage of disruption.  
\end{abstract}

\section{Introduction}\label{sec:intro}
Stars are born in dense molecular clouds, and those that remain gravitationally bound appear as star
clusters \citep{lad03}. Stars inside a cluster interact with each other via two-body relaxation.
As a consequence, massive members slow down and sink to the cluster center, where low-mass stars
speed up occupying a larger column progressively and eventually escape. The so-called ``mass segregation''
is often observed in star clusters \citep{hil98,pan13,tan18}. In the meanwhile, Galactic potential 
perturbs star clusters leading to the formation of tidal structures. For example, giant tidal tails
have been found in the isolated halo globular cluster Palomar 5 \citep{ode01,ode03}. 

The Galactic disk is abundant in stars, spiral arms, and giant molecular clouds. Therefore, star
clusters located in the disk are subjected to disturbance, such as disk shock, spiral arm passage,
molecular cloud encounters, etc. \citep{spi58, kru12}. The typical survival timescale of open
clusters in the Galactic disk is about 200~Myr \citep{bon06, yan13}. Open clusters much older than
the survival timescale must have their shape distorted, and structure loosened, leading to inevitable 
disruption. The disintegrated open clusters become moving groups and then supply field stars. 
A small portion of these stellar group remnants can be identified by convergent-point method 
\citep{bos08} in the solar neighborhood
\citep[such as the TW~Hydrae association, AB doradus moving group and more;][]{zuc04}.

However, detection of tidally disrupted substructures of open clusters is painstaking. The low
number density of a tidal tail may cause its members to be buried in the dense foreground and 
background field stars. Accurate kinematic data are essential to recover such substructures. {\it Gaia}
data revolutionize this study by providing high precision proper motion (PM) and parallax ($\varpi$) 
for nearby open clusters \citep[e.g.,][]{can18}. Based on the 3D motions from {\it Gaia}~TGAS catalog,
\citet{oh17} identified more than 4555 moving groups in the solar neighborhood. However, only 61 of
them have members more than 5, among them, ten are related to known associations. The second data
release (DR2) of {\it Gaia} \citep{bro18} with a higher accuracy on kinematic data, makes it possible
to directly reveal tidal tails, e.g., in the nearest open cluster, Hyades \citep{ros18,mei18}. 

The open cluster Coma Berenices (Melotte\,111, hereafter Coma Ber) with an age around 800~Myr
\citep{tan18} is the second nearest \citep[86.7~pc;][]{tan18} star cluster 
to the Sun. However, with a large sky coverage and an average PM with no significant difference from 
that of the field stars, Coma Ber has gotten less attention compared to other nearby clusters. The earliest 
studies on Coma Ber can be dated back to \citet{mel15} and \citet{tru38}. Later on, \citet{ode98}
used the {\it Hipparcos} and {\it Tycho} plus the ACT Reference Catalog \citep{urb98}, to perform,
for the first time, a detailed study with parallax information of Coma Ber, and found a core-halo
structure with the major axis parallel to the direction of the Galactic orbital motion of the cluster.
They also detected a group of stars with a tangential distance $>$ 10~pc from the cluster center, which 
they called the ``moving group'' of Coma Ber. By studying the luminosity function, they found more 
faint stars in the moving group than in the central cluster, therefore concluded that Coma Ber was
under the process of dissolution. Similar conclusion was given in the later studies by \citet{cas06},
\citet{kra07}, and \citet{tan18}.

Using the data from {\it Gaia} DR\,2, we explore the neighborhood of Coma Ber and search for tidal 
tail substructures. In Section~\ref{sec:data}, we introduce the quality and limitation of the 
{\it Gaia} DR\,2 data, and explain our input data-set for structure identification. We then present 
the algorithm, \textsc{StarGO}, which is used to identify structures. The results are shown in 
Section~\ref{sec:result}. The dynamical status of Coma Ber and the nearby associated structures, 
Group-X is discussed in Section~\ref{sec:Discussion}. Finally, we provide a brief summary in 
Section~\ref{sec:Summary}.

\section{Data and Analysis}\label{sec:data}

\subsection{Gaia DR\,2 Photometry and Kinematics}\label{sec:gaia}

As the successor to the {\it Hipparcos} telescope, {\it Gaia} is an on-going space mission carried
out by the European Space Agency, aiming at providing a detailed three dimensional map on positions
and space motions of about one billion stars in our Galaxy and beyond. The DR2 of {\it Gaia} \citep{bro18}
has provided approximately 1.7 billion sources with the celestial positions (R.A. and decl.), and the
$G$~band (330--1050 nm) photometry, whose magnitude ranges from $\sim$ 3 to 21~mag. Additionally, two
other broad-band photometry, $G_{\rm BP}$ (330--680~nm) and $G_{\rm RP}$ (630--1050~nm) are available
for around 1.4 billion sources for the very first time.

The astrometry solutions in DR2 have much higher precisions compared to DR1 because they
are derived from data collected in a longer time span (22 August 2014 to 23 May 2016) of the nominal 
mission lifetime. Parallaxes and PMs ($\mu_\alpha \cos\delta, \mu_\delta$) are available
for about 1.3 billion sources. The median uncertainty of $\varpi$ without considering the systematic
errors is $\sim$0.04~mas for bright sources with $G$\,$<$\,14~mag, 0.1~mas for $G$\,$\approx$\,17~mag, 
and $\sim$0.7~mas for $G$\,$\approx$\,20~mag. The corresponding uncertainties of PMs (without 
considering the systematic errors) for these sources are 0.05, 0.2 and 1.2\,mas~yr$^{-1}$, respectively 
\citep{lin18}.

The RV spectrometer on the {\it Gaia} telescope collects spectra with a medium resolution (R$\sim$11700) 
in the wavelength range of 845--872 nm, centering at the calcium triplet region. However, the radial 
velocity (RV) in DR2 is only available for bright sources ($\sim$\,7.2~million stars), with a typical 
uncertainty less than 2\,$\rm km~s^{-1}$ \citep{bro18}. 


\subsection{Input Data}\label{sec:input}

The N-body simulation of \citet{ern11} has shown that tidal tails of an open cluster can be as long
as 800\,pc, with a core region less than 100\,pc (Figure~6 in \citet{ern11}). However, this is likely 
an upper limit since the simulation only considered Galactic potential as the tidal source, without 
external perturbations such as molecular cloud interaction, disk shocking, or spiral arm passages. 
Conceivably, only the most inner part of tidal tails close to the cluster center can be preserved.
In this work, we search for tidal tails around the Coma Ber within a radius of 85 pc of its center.
Here, we take the distance of Coma Ber as 86.7 pc \citep{tan18}, and the equatorial coordinates of 
its center as (R.A.=$12^{\rm h} 25^{\rm m}$, decl.=$26\degr 06^{\prime}$, J2000) from \citet{dia14}. 
From the above, the Cartesian Galactocentric coordinates of the cluster center is 
($X, Y, Z$) = ($-8304.9$, $-5.9$, $+112.1$)~pc.

\begin{figure}[tb!]\centering
	\includegraphics[width=\columnwidth]{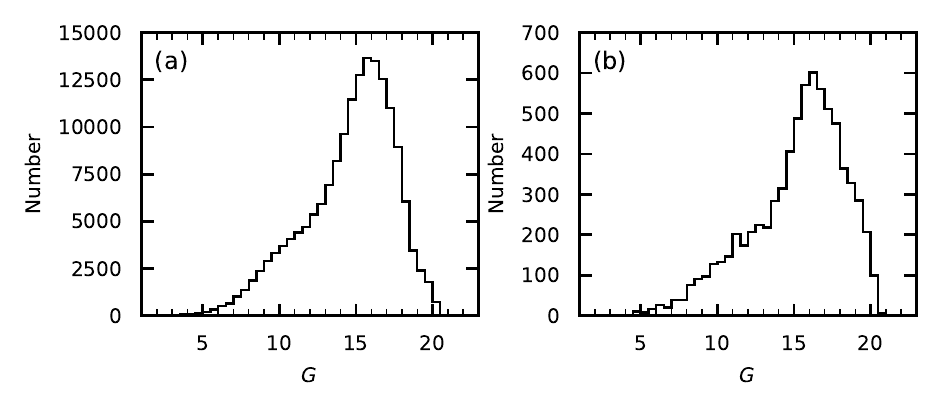}
	\caption{ 
	 Number of {\it Gaia} DR\,2 stars in $G$ band of  Sample~I is shown in (a), 
	and Sample~II  shown in (b).
	} 
  \label{fig:ghist}
\end{figure}

\begin{figure}[tb!]\centering
	\includegraphics[width=\columnwidth]{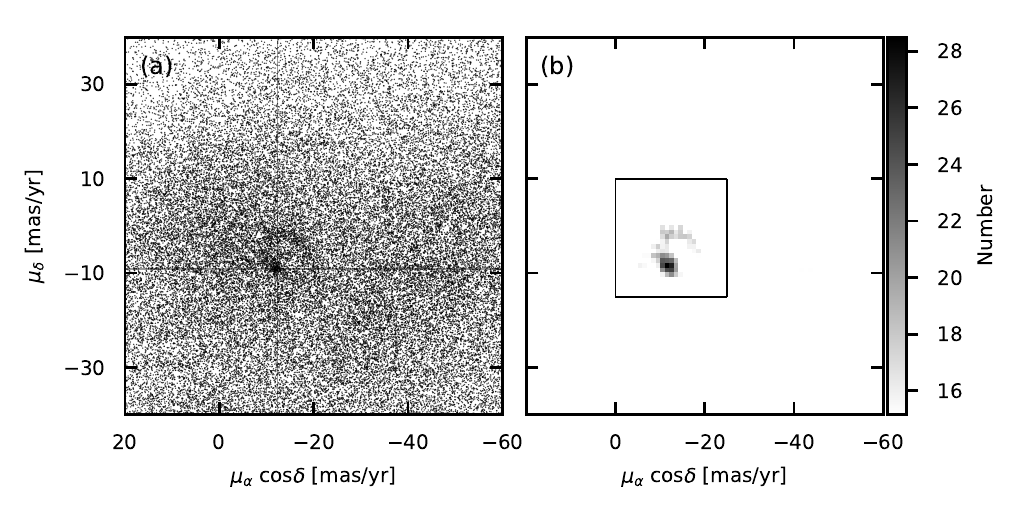}
	\caption{ 
	(a)~ Proper motion vector plot for sample~I around Coma Ber, 
	    which generates an over-density as the cross indicated. 
	(b)~ 2D density map for sample~I. Each bin is smoothed by 
	    neighboring 8 bins and here only bins with a number count $>$ 12.8 
	    (5$\sigma$, where $\sigma$ is the standard deviation of all bins) 
	    are shown. The black box indicates the PM ranges of Sample~II. 
	} 
  \label{fig:2dhist}
\end{figure}

To exclude possible artifacts in {\it Gaia} DR\,2, we apply the astrometric quality cuts:  
$\varpi/ \Delta \varpi>10$, and the ``astrometric excess noise'' $<1$~mas, suggested by \citet{lin18} 
(in their Appendix C). Furthermore, we only use RV data with errors  $\rm <1\,km~s^{-1}$. The final
``cleaned'' sample contains 152,739 sources which we call Sample~I. This sample has the $G$ magnitudes
ranging from $\sim$3.2~mag to $\sim$20.7~mag, with the distribution function turning around, i.e., being 
significantly incomplete, beyond $\sim$16~mag, as shown in Figure~\ref{fig:ghist}~(a).
About 25 percent of the sources in Sample~I have available RV. Figure~\ref{fig:2dhist}~(a) shows the PM 
vector plot of Sample~I around Coma Ber. Interestingly, an additional belt-like over-density is seen. In 
Figure~\ref{fig:2dhist}~(b), the belt-like feature reveals itself robustly on a 2D density map that 
only shows bins with an over-density $>5\sigma$. To include this belt-like structure and Coma Ber, 
we apply a cut in PMs, $\mu_\alpha \cos\delta$ between $-25$ and $0$~mas~yr$^{-1}$ and $\mu_\delta$
between $-15$ and $10$~mas~yr$^{-1}$ (the black box in Figure~\ref{fig:2dhist}~(b)). This reduces
the number of stars to 5,494, which is called Sample~II.
This smaple is very similar to Sample~I with magnitudes ranging from $\sim$3.0~mag to $\sim$20.8~mag, 
and complete near $\sim$16.5~mag (see Figure~\ref{fig:ghist}~(b)).

In this study, we use 5D parameters of stars in Sample~II 
(R.A., decl., $\varpi$, $\mu_\alpha \cos\delta$, and $\mu_\delta$)
from {\it Gaia} DR2. Since only a fraction of stars ($\sim$23 percent) have RV measurements, RV is used
supplementarily. Stars in Sample~II are all located within 175\,pc from the Sun, which makes the most 
probable distance of each star very close to the inverse of $\varpi$ \citep{bai18}. Therefore, adopting
1/$\varpi$ as the distance, we compute for each source the Galactocentric Cartesian coordinates 
($X, Y, Z$). The transformation is performed by using the Python \texttt{Astropy} package \citep{ast13,ast18}.
Assumptions used in the conversion, such as the solar position and its Galactic velocities are listed in 
Appendix~A.

\subsection{Structure Identification with \textsc{StarGO}}\label{sec:identify}

We use a novel cluster finding method \textsc{StarGO} \citep{yuan18} to search for tail structures
around Coma Ber. Our method is built with self-organizing map (SOM), which is one of the well-established
unsupervised learning algorithms based on artificial neural network. The goal of SOM is to create a network
that stores the information in such a way that the topological structures of the input data are preserved. 
Based on that, structures having grouping signature in the input space will be more easy to be identified.

An open cluster is a group of stars formed from the same giant molecular cloud. Before the member stars are
fully dissolved, they are clustered in the phase space. In this work, we apply \textsc{StarGO} to the stellar 
sample around Coma Ber in the 5D space of ($X, Y, Z$, $\mu_\alpha \cos\delta, \mu_\delta$). The unsupervised 
learning of the neural network is performed as follows. 

\begin{enumerate}

\item We generate a 2D network with 100$\times$100 neurons located at different grid points. The distances
between neurons are defined as the Euclidean distances on the 2D map.  Each neuron is assigned with a 5D
weight vector which has the same dimension as the input space.

\item From the sample, we pick one star and find the neuron with the closest weight vector to its input 
vector, which is defined as the best matching unit (BMU). Each neuron updates its weight vector 
according to its distance to the BMU. The learning algorithm enforces the neurons, that located further
from the BMU on the map, get less affected by the input star.

\item Stars are fed to the neural network one by one, where all the neurons keep updating their weight
vectors from the previous values. One iteration is complete after the neurons learn the behavior from
all the stars of the sample once. The whole learning process is iterated by 400 times when the weight 
vectors reach convergence. We associate all the stars from our sample to the final map with their BMUs.
\end{enumerate}

\begin{figure*}[tb!]\centering
	\includegraphics[width=.9\textwidth]{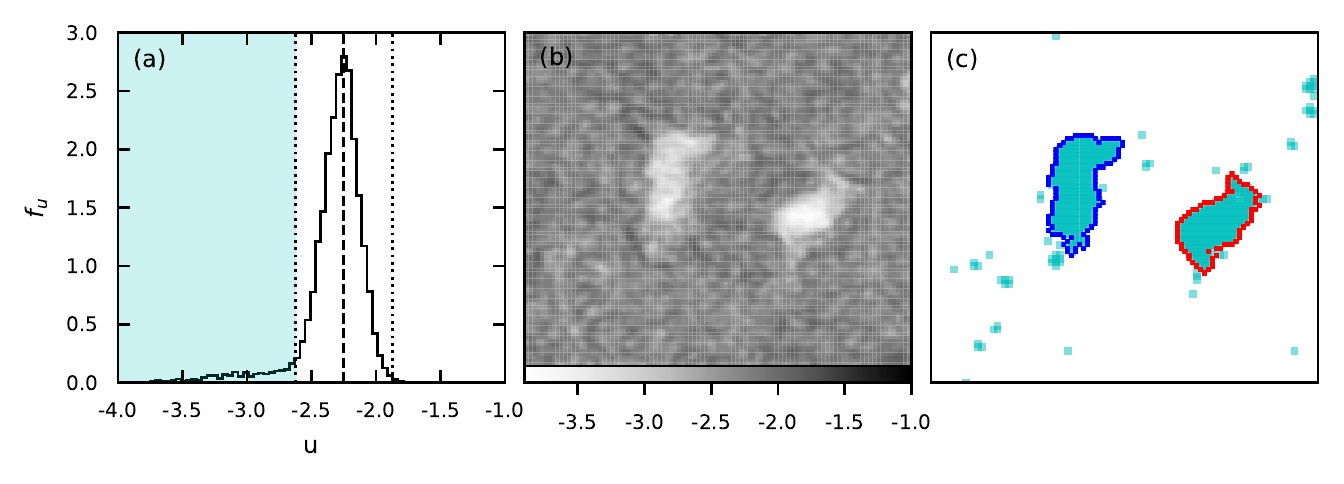}
	\caption{
	Group identified by applying \textsc{StarGO} to the data sample in the 
	($X, Y, Z$, $\mu_\alpha \cos\delta, \mu_\delta$) space. 
	(a) histogram shows the distribution of {\it u}. The dashed lines
	    denotes the peak position of its distribution $u_{\mathrm{peak}}$ and the dotted lines
	    show $u_{\mathrm{peak}-3\sigma}$ and $u_{\mathrm{peak}+3\sigma}$, respectively. 
	    The left tail with $u<u_{\mathrm{peak}-3\sigma}$ is highlighted in cyan.
	(b) 2D neural map resulting from SOM, where the u value between adjacent neurons is represented
	    by the gray-scale. 
	(c) is the same as (b) where the neurons with $u<u_{\mathrm{peak}-3\sigma}$ are colored in
	    cyan. Contours of two dominant neuron groups are traced with different colors (red: Coma Ber; blue: group-X).
	}
  \label{fig:som}
\end{figure*}

The differences in weight vectors between adjacent neurons are denoted by a matrix, $u$. The trained neural 
network can be visualized by the gray scale map of $u$ (as illustrated in Figure~\ref{fig:som}~(b)). 
The lighter gray shaded neurons have smaller $u$, which means that their weight vectors are more similar
to their neighbors compared to the darker shaded ones. Thus the stars associated with the former are closer 
in the input space. As we can see clearly from Figure~\ref{fig:som}~(b), the lighter shaded neurons form two 
distinctive patches. These neurons have significantly lower values of $u$ than the others, which are supposed 
to be associated with the stars clustered in the input space. Consistently, we see an extended tail on the 
left side of the distribution of $u$ in Figure~\ref{fig:som}~(a), indicating that there are some neurons
with significant low values of u. We take the following steps to select these neurons. Firstly, we find the
peak position $u_{\mathrm{peak}}$ and the 99.85th percentile of the distribution $u_{99.85\%}$, which are 
denoted by the dotted lines respectively in Figure~\ref{fig:som}~(a). The quantity $u_{99.85\%}-u_{\mathrm{peak}}$ 
is equivalent to 3$\sigma$ of a normal distribution, which is denoted as $\Delta_{3\sigma}$. We then define the 
left tail of this distribution by $u_{\mathrm{peak-3\sigma}}~<~u_{\mathrm{peak}}-\Delta_{3\sigma}$, shown as 
the cyan shaded region in Figure~\ref{fig:som}~(a). We mark all the neurons with $u\leq u_{\mathrm{peak-3\sigma}}$
by cyan in Figure~\ref{fig:som}~(c), and color the two most major groups of 197 and 218 stars in red and blue, 
respectively. These two groups correspond to Coma Ber and its nearby group (here and after group-X), which will
be discussed in detail in Section~\ref{sec:result}. 
\subsection{Uncertainty and Contamination}\label{sec:obs_uncert}

The observational uncertainties are taken into account by using the trained 2D neuron map. 
We first compute the difference between the input vector of each star and its BMUs, denoted by
$u_{\rm{vw}}$. The maximum difference from the training sample is quantified as $u_{\rm{vw,max}}$, 
which denotes the limit of the neuron map. For each member candidate of the identified group, we 
create 1000 realizations which follow the Gaussian distribution in each dimension of 
($\varpi$, $\mu_{\alpha} \cos\delta$, $\mu_{\delta}$) with the observed value as the mean and the 
uncertainty as the covariance. We find BMU for each mock realization star from the trained map. We 
then calculate the difference between the input vector of the mock star and its associated BMU. For 
stars with large observational errors, the mock realizations may have $u_{\rm{vw}}$ larger than 
$u_{\rm{vw,max}}$. We remove those stars since they cannot be tightly associated with the trained 
neuron map. For each candidate members, more than 900 out of 1000 of its mock realization attach 
to the corresponding group. In another word, the probability to recover member candidates is more 
than 90\%, considering observed uncertainties.

We then evaluate the contamination fraction from the smooth Galactic disk background using the mock 
{\it Gaia} DR\,2 catalog of \citet{ryb18}, which is based on the \texttt{Galaxia/Besan\c{c}on} models
\citep{galaxia,besancon}. The same cut on $\varpi$ and PM as per Section.~\ref{sec:input} were 
exerted on the mock catalog in the same volume of the sky (within 85 pc around the center of Coma 
Ber). There are 3,494 and 592 mock stars within the sky area of the two identified clusters, respectively.
Following the recipe described in the previous section, we attach each of these mock stars with
the trained map, and find 20 stars associated with Coma Ber and 12 stars associated with group-X.
Therefore, the mock smooth background gives 20/3,494 = 0.6$\%$ of field stars in the region of Coma Ber,
and 12/592 = 2$\%$ in the region of group-X. Considering the observed stars within the sky area of the two
identified clusters (1,794 and 658), the contamination fraction estimated from the mock data is 5\%--6\% 
(Coma Ber: $1,794\times0.6\%/197$; group-X: $658\times2\%/218$).

We also verify the choice of the searching volume for cluster identification by enlarging the radius 
to 100\,pc and 150\,pc. In either case, similar member lists of the targeted open clusters are 
obtained. We do not find significantly extended structures beyond 85\,pc for these two clusters. 
Therefore in this paper, we only present the results within the fiducial volume.

\section{Identified Structures}\label{sec:result}
\subsection{Coma Ber and its Tidal Tail}\label{sec:coma}

\begin{figure*}[tbh!]\centering
	\includegraphics[width=1.\textwidth]{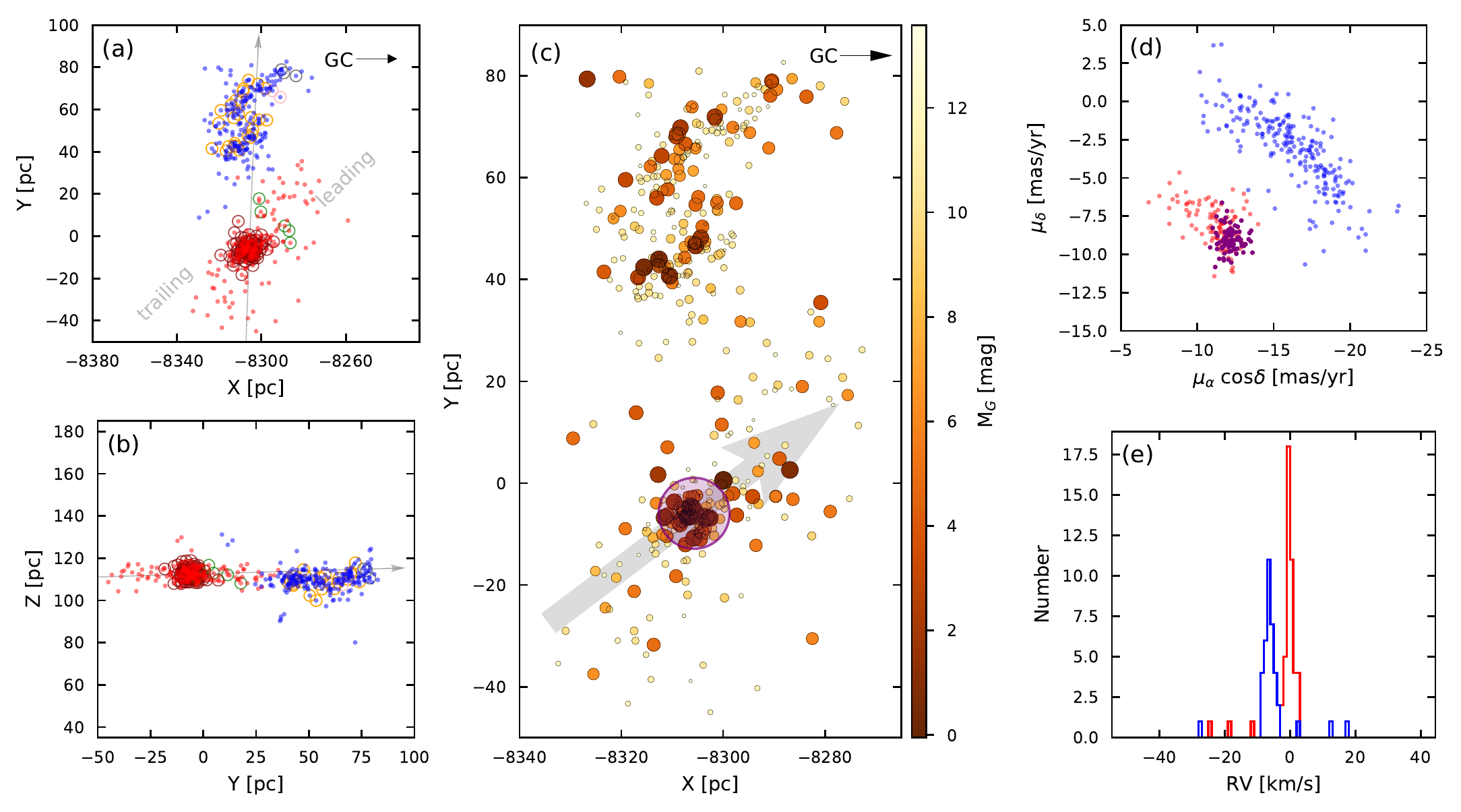}
	\caption{
	Kinematic and spatial distribution of member candidates
	(Coma Ber: red; group-X: blue). Members from previous studies are in open circles 
	(\citet{ode98}: green; \citet{tan18}: dark red; \citet{oh17}'s group 10: orange; 
	\citet{oh17}'s group 81: grey; \citet{oh17}'s group 1805: pink).
	(a) and (b) show the 3D spatial position of member candidates in 
	    Galactocentric Cartesian $X,Y,Z$ coordinates. The grey arrow in (a) and (b) is the  
	    orbit motion of Coma Ber.
	(c) is the same as (a), but with the symbols color coded and sizes adjusted to reflect
	    the M$_G$~magnitude. Bright stars are bigger and with darker color. The purple circle
	    shows the region within the tidal radius (6.9~pc, see Section~\ref{sec:radii}) of Coma Ber. 
	    The big grey arrow is the projected mean velocity vector of Coma Ber relative to LSR. 
	(d) shows the proper motion vector plot with purple dots as the Come Ber member candidates 
	    within the tidal radius.
	(e) shows the RV histogram. 
}
  \label{fig:sp_vol}
\end{figure*}

In Figure~\ref{fig:sp_vol}~(a) and (b), we show the 3D spatial distributions of Coma Ber 
member candidates in Table~\ref{tab:coma}, with the running number (column 1), the astrometric
(position, $\varpi$ and PM) and photometric (magnitude) information from {\it Gaia} DR\,2 listed
in column 2 to 12. The cluster is already known to be elongated along the Z axis \citet{tan18} 
with most of the extended members fainter than the brightness limit of Gaia DR\,2. In this study, 
we find the elongation on the X--Y plane, leading to an overall morphology of an ellipsoid. 

The differential gravitational force of the Milky Way, i.e., the Galactic tide, acts to stretch
objects such as open clusters \citep{mar17}. The Galactic tide is extensive along the radial
direction, and compressive along the vertical direction of the plane \citep{spi87}, thereby 
re-shaping open clusters into an ellipsoid. Stars in an ellipsoid have different Galactocentric
distances. Due to the differential rotation, stars closer to the Galactic center rotate faster.

Therefore, one part of the ellipsoid is leading while the other is trailing. This is what is seen
in the morphology of Coma Ber, shown in Figure~\ref{fig:sp_vol}~(a) and (c) , and is discovered 
for the first time for this star cluster.

We show the orbit motion of Coma Ber with a grey arrow in panels (a) and (b). The orbit is 
computed base on the average position of Coma Ber (Section~\ref{sec:input}) and its median velocity,
by the Python \texttt{galpy} package \citep{bov15} with the Galactic potential, 
``\texttt{MWPotential2014}\footnote{\texttt{MWPotential2014} is a Galactic potential model made of 
three components, the bulge, disk, and halo. Parameters of the model were fitted to 
published dynamical data of the Milky Way.}''.
Orbits are integrated 0.5\,Myr backward and 1\,Myr forward in time.

Figure~\ref{fig:sp_vol}~(c) is the same as Figure~\ref{fig:sp_vol}~(a) but with the color coding and
symbol size scaled to the absolute $G$ magnitude (M$_{G}$). As can be seen, brighter stars are 
concentrated toward the center, which confirms the mass segregation previously known in Coma Ber
\citep{ode98,cas06,kra07,tan18}. Compared to the local standard of rest \citep[LSR;][]{ker86}
the extension of tidal tails are aligned with the major axis of the ellipsoid of Coma Ber indicated 
by the grey big arrow in Figure~\ref{fig:sp_vol}~(c), which is the projection of mean velocity
vector of Coma Ber (U, V, W)=(8.6, 226.6, 6.8)~$\rm~km~s^{-1}$ relative to 
LSR: (U, V$-220$~$\rm~km~s^{-1}$, W)=(8.6, 6.6, 6.8)~$\rm~km~s^{-1}$. 

A recent work by \citet{tan18} has carried out a comprehensive study on Coma Ber within a 5$\degr$
radius sky area for stars from the bright end ($J$ $\sim$ 3~mag $\sim$2.3~$M_\sun$) down to the brown
dwarf regime ($J$ $\sim$ 17.5~mag $\sim$0.06~$M_\sun$). For member candidates outside the 5$\degr$
radius, they had collected from other studies \citep{tru38,cas06,kra07,mer08,mel12,van17} if the
candidate also satisfied \citet{tan18}'s selection criteria. Among their 154 bright members, our
candidate list (red dots in Figure~\ref{fig:sp_vol}) recovers 82 stars ( dark red open circles in
Figure~\ref{fig:sp_vol}~(a) and (b)), which are located in the cluster center. The others might
be field star contamination since \citet{tan18} applied a large radius cut (17~mas~yr$^{-1}$) on PM
selection (see their Figure~4) compared to the very concentrated PM of our member candidates (see 
Figure~\ref{fig:sp_vol}~(d)) with standard deviations of
$\sigma_{\mu_\alpha \cos\delta}=$1.2~mas~yr$^{-1}$ and $\sigma_{\mu_\delta}=$ 1.2~mas~yr$^{-1}$. 

Five stars identified by \citet{ode98}, HD~114400, BD+21~2514, BD+38~2436, HD~116706, and BD+26~2461,
were called as Coma Ber's moving group. They are recovered by the \textsc{StarGO} algorithm as
Coma Ber's member candidates (see green open circles in Figure~\ref{fig:sp_vol}~(a) and (b)). These
five stars are indeed located further away from the main cluster, belonging to the tidal tail
structures.

\subsection{Group-X}

Although Coma Ber and the group-X contain similar numbers of stars, they are distinct from each other
in the PM vector diagram (Figure~\ref{fig:sp_vol}~(d)). The range of the PM of group-X is wider than
that of Coma Ber, with standard deviations of $\sigma_{\mu_\alpha \cos\delta}$ = 2.6~mas~yr$^{-1}$
and $\sigma_{\mu_\delta}$ = 2.5~mas~yr$^{-1}$. Besides, we also present the RV distributions
for 47 member candidates in Coma Ber and 38 in group-X, which have reliable RV measurements
(Figure~\ref{fig:sp_vol}~(e)). The distinction of two separate peaks at $-0.5$~km~s$^{-1}$ (Coma Ber)
and $-7.0$~km~s$^{-1}$ (group-X) further confirms the fact that Coma Ber and group-X are two
dynamically different systems. The narrow distributions of RV of both groups suggest that their
members are kinematically related. 

Despite the proximity of group-X, among the 218 member candidates we identify, 27 were first
discovered by \citet{oh17} via the $\varpi$ and PMs from {\it Gaia} DR1~TGAS data. These 27 stars
were members of group~10 in \citet{oh17}'s Table~1 (hereafter Oh10), and were confirmed as a newly
found moving group in \citet{fah18}. We plot Oh10's 27 stars as  orange open circles in 
Figure~\ref{fig:sp_vol}~(a) and (b). They are all bright stars and located in two dense patches,
tracing out the shape of the two clumps (most of dark and large circles in group-X in
Figure~\ref{fig:sp_vol}~(c)). In addition to Oh10, 3 member candidates of group-X belong to group~81
in \citet{oh17} (hereafter Oh81, grey open circles), and two belong to their group~1805 (hereafter
Oh1805, pink open circles). 

Besides Oh10, Oh81, and Oh1805, the remaining 186 stars are dynamically associated as a
group in Figure~\ref{fig:sp_vol}~(d) and (e). The extended distribution of PMs of group-X may
reflect the disrupted nature of a moving group for which all members might share the same origin.
The member candidates for group-X are listed in Table~\ref{tab:groupx}, with columns 2 to 12
presenting the astrometric and photometric data from {\it Gaia} DR\,2 (the same format as 
Table~\ref{tab:coma}).


\startlongtable
\begin{deluxetable*}{ccc rrRR RRRR rc}
\tablecaption{Coma Ber Member Candidates\label{tab:coma}
             }
\tabletypesize{\scriptsize}

\tablehead{
	 \colhead{No.}		                     & \colhead{R.A.}                & \colhead{Decl.}                   &
	 \colhead{Plx}                           & \colhead{Plxerr} 	         & \colhead{$\mu_\alpha \cos\delta$} &
	 \colhead{$\Delta\mu_\alpha \cos\delta$} & \colhead{$\mu\delta$}         & \colhead{$\Delta\mu\delta$}       &
     \colhead{RV}                            & \colhead{RVerr}               & \colhead{$G$}                     &    
     \colhead{flag$^\dagger$}                                                                                    \\
	 \colhead{}                              & \multicolumn{2}{c}{(J2015.5 deg)}                                 &
	 \multicolumn{2}{c}{(mas)}               &
	 \multicolumn{2}{c}{(mas~yr$^{-1}$)}     &
	 \multicolumn{2}{c}{(mas~yr$^{-1}$)}     & \multicolumn{2}{c}{(km~s$^{-1}$)}                                 &
     \colhead{(mag)}                         &  \colhead{}
	 }
\colnumbers
\startdata
1  & 183.103668 & 27.380058 & 11.67 & 0.05 & -12.19 & 0.08 & -9.41  & 0.05 & 0.15    & 0.27    & 7.97  & b \\
2  & 183.178196 & 25.228414 & 11.31 & 0.12 & -11.75 & 0.17 & -7.85  & 0.12 & \nodata & \nodata & 16.52 & b \\
3  & 183.221756 & 26.250356 & 11.59 & 0.04 & -12.07 & 0.06 & -9.52  & 0.04 & 0.59    & 1.01    & 11.10 & b \\
4  & 183.881094 & 25.066959 & 11.32 & 0.11 & -12.13 & 0.17 & -9.13  & 0.14 & \nodata & \nodata & 16.04 & b \\
5  & 184.034840 & 25.760327 & 11.72 & 0.05 & -12.23 & 0.08 & -10.53 & 0.06 & 2.06    & 0.37    & 7.98  & b \\
78 & 164.212295 & 17.330554 & 9.79  & 0.17 & -10.73 & 0.29 & -8.26  & 0.30 & \nodata & \nodata & 17.18 & t \\
79 & 165.493305 & 20.033749 & 10.28 & 0.07 & -11.03 & 0.12 & -8.92  & 0.16 & \nodata & \nodata & 15.04 & t \\
80 & 166.639954 & 14.234920 & 10.18 & 0.04 & -10.95 & 0.07 & -9.10  & 0.06 & 1.77    & 0.66    & 11.38 & t \\
81 & 168.564244 & 9.241932  & 10.27 & 0.12 & -10.87 & 0.18 & -7.36  & 0.17 & \nodata & \nodata & 16.86 & t \\
82 & 170.489998 & 23.773542 & 11.74 & 0.13 & -12.24 & 0.22 & -11.05 & 0.20 & \nodata & \nodata & 16.99 & t 
\enddata
\tablecomments{
    Entries are sorted according to column 2, R.A.. 
    This table is available in its entirety in a machine-readable form in the online journal.
    Here we only show the first five member candidates in both the bound (within tidal radius) and the tail regions.\\
    $^\dagger$ b: ``bound'' member candidates within tidal radius (see Section~\ref{sec:radii}), t: ``tail'' member candidates. 
    }
\end{deluxetable*} 


\startlongtable
\begin{deluxetable*}{ccc rrRR RRRR r}
\tablecaption{Group-X Member Candidates\label{tab:groupx}
             }
\tabletypesize{\scriptsize}

\tablehead{
	 \colhead{No.}			                 & \colhead{R.A.}                   & \colhead{Decl.}                   &
	 \colhead{Plx}                           & \colhead{Plxerr} 	            & \colhead{$\mu_\alpha \cos\delta$} &
	 \colhead{$\Delta\mu_\alpha \cos\delta$} & \colhead{$\mu\delta$}            & \colhead{$\Delta\mu\delta$}       &
     \colhead{RV}                            & \colhead{RVerr}                  & \colhead{$G$}                     \\  
	 \colhead{}                              & \multicolumn{2}{c}{(J2015.5 deg)}                                     &
	 \multicolumn{2}{c}{(mas)}               & 
	 \multicolumn{2}{c}{(mas~yr$^{-1}$)}     &
	 \multicolumn{2}{c}{(mas~yr$^{-1}$)}     & 
	 \multicolumn{2}{c}{(km~s$^{-1}$)}       & \colhead{(mag)}
	 }
\colnumbers
\startdata
1  & 178.948766 & 39.073363 & 9.20  & 0.04 & -21.02 & 0.06 & -9.76 & 0.05 & 2.88    & 0.99    & 10.08 \\
2  & 181.613107 & 40.057117 & 9.69  & 0.06 & -18.27 & 0.06 & -9.78 & 0.08 & \nodata & \nodata & 15.18 \\
3  & 187.799135 & 38.782755 & 9.64  & 0.05 & -16.22 & 0.06 & -2.35 & 0.05 & 17.85   & 0.24    & 8.99 \\
4  & 191.762024 & 50.574560 & 11.07 & 0.15 & -19.93 & 0.19 & -8.31 & 0.21 & \nodata & \nodata & 17.87 \\
5  & 193.693214 & 49.169553 & 11.65 & 0.06 & -20.77 & 0.07 & -8.33 & 0.08 & \nodata & \nodata & 16.03 \\
6  & 195.253545 & 46.072319 & 11.73 & 0.05 & -20.54 & 0.06 & -8.89 & 0.06 & \nodata & \nodata & 14.77 \\
7  & 195.729678 & 55.236239 & 10.30 & 0.08 & -18.82 & 0.11 & -7.93 & 0.12 & \nodata & \nodata & 16.88 \\
8  & 195.819464 & 57.315206 & 10.54 & 0.03 & -17.37 & 0.04 & -8.33 & 0.04 & -3.99   & 0.90    & 8.97 \\
9  & 197.861213 & 60.205489 & 12.60 & 0.07 & -21.26 & 0.10 & -7.17 & 0.10 & \nodata & \nodata & 15.58 \\
10 & 198.797593 & 50.390444 & 10.09 & 0.07 & -15.08 & 0.09 & -7.00 & 0.09 & \nodata & \nodata & 15.85 
\enddata
\tablecomments{
    Entries are sorted according to column 2, R.A.. 
    This table is available in its entirety in a machine-readable form in the online journal.
    Here we only show the first ten candidates.
    }
\end{deluxetable*} 

\section{Discussion}\label{sec:Discussion}
\subsection{Cluster Ages}\label{sec:age}

\begin{figure*}[tb!]\centering
	\includegraphics[width=.9\textwidth]{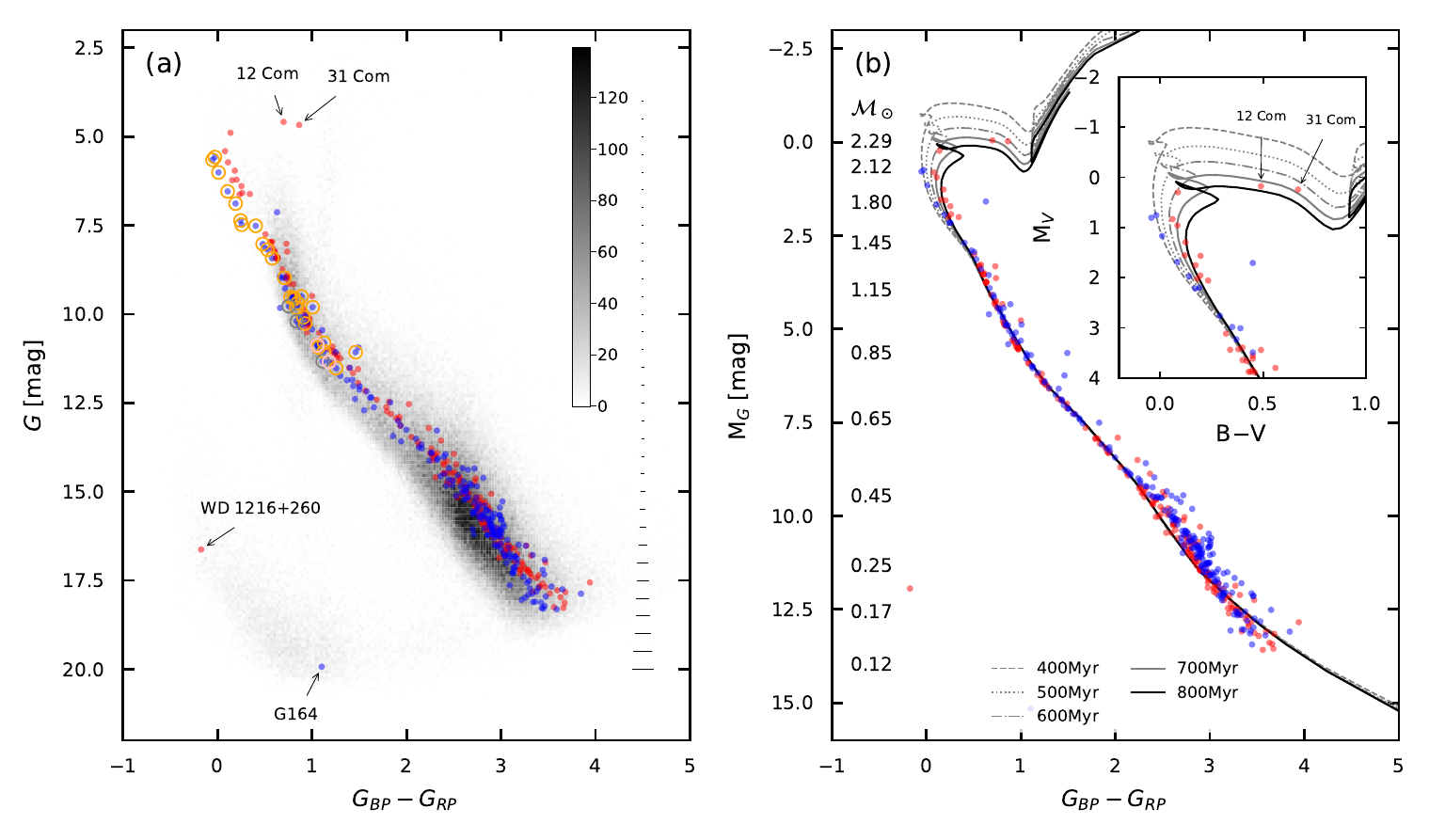}
	\caption{
	    The color-magnitude diagram of 
	    (a) {\it Gaia} apparent $G$ magnitude, and (b) the absolute magnitude M$_{G}$ (adopting {\it Gaia} parallax). 
	        Symbol are the same as in Figure~\ref{fig:sp_vol}~(a).
	        Objects of Sample-I are shown as a density map in (a). Typical errors in color
	        $G_{\rm BP}-G_{\rm RP}$ are presented as horizontal bars to the right. 
	    (b) PARSEC isochrones of 400, 500, 600, 700, and 800~Myr are over-plotted, with zero
	        extinction corrected and solar metallicity assumed. Stellar masses, per the 800~Myr
	        isochrone are indicated. 
	    The inset in (b) shows the MS turn-off with $B$ and $V$ photometry. 
	    }
  \label{fig:cmd}
\end{figure*}

We plot the color-magnitude diagram (CMD) for both groups (Coma Ber: red; group-X: blue) in 
Figure~\ref{fig:cmd} with stars from Sample~I plotted as a density map in the background in
Figure~\ref{fig:cmd}~(a). There are 25 member candidates of Coma Ber and 18 of group-X having
spectra from the Large Sky Area Multi-Object Fiber Spectroscopy Telescope (LAMOST) DR\,4, with
S/N$>$40 in $g$~band \citep{wu14a, wu14b}. Both have nearly solar metallicity (Coma Ber to be 
[Fe/H] $\sim-0.02$ and group-X $\sim0.08$), hence we adopt [Fe/H]=0 and zero extinction 
\citep[in the solar neighborhood,][]{cas06,tan18} for the isochrone, and correct each star for
its distance and obtain the CMD in M$_{G}$ (Figure~\ref{fig:cmd}~(b)).

Recent studies \citep{eva18,wei18,maz18} have provided three different sets of the filter response
curves for {\it Gaia} in $G$, $G_{\rm BP}$, and $G_{\rm RP}$ bands, thus leading to three different
sets of PARSEC isochrones \citep{bre12,che14,tan14,che15}, with a large offset ($\leq0.3$~mag) in 
the $G_{\rm BP}$ band for bright stars. Here we adopt the PARSEC isochrones with the sensitivity
curves provided by \citet{maz18} (best match to our data) and plot them in Figure~\ref{fig:cmd}~(b).
For consistency check, we also present the CMD for bright stars in $B$ and $V$ photometry taken from
Tycho-2 catalogue \citep{hog00} in the inset of Figure~\ref{fig:cmd}~(b).

The age of Coma Ber is determined by the stars near the main sequence turn-off point, 
and two stars, 12 Com and 31 Com, just evolving off the main sequence, using the PARSEC isochrone 
set. Given that 12~Com is a known binary \citep{gri11}, hence brighter than a single-star isochrone of 
the actual age, a range of 700--800 Myr PARSEC isochrone provides the overall best fit. This is consistent 
with the result of \citet{tan18}.

The member, WD~1216+260 (in Figure~\ref{fig:cmd}~(a)), is spectroscopically confirmed as a DA white dwarf
\citep{dob09, gir11}. According to the cooling time of the white dwarf estimated by \citet{gir11}, 
WD~1216+260 must be older than 400\,Myr, which sets a lower limit for the age of Coma Ber. This agrees 
with the age range quoted in the literature. 

In contrast, no turn-off stars are present in the CMD of group-X, indicating a younger age. 
The upper MS of group-X is bluer than Coma Ber, and located very close to the isochrone of 400\,Myr
in $G$ versus $G_{\rm BP}-G_{\rm RP}$ diagram (Figure~\ref{fig:cmd}~(b)). Furthermore, in the $B$
versus $B-V$ diagram, the bright stars clearly line up along the 400\,Myr isochrone (inset in
Figure~\ref{fig:cmd}~(b)). We adopt an age 400\,Myr for group-X from isochrone fitting by eye via
$B$ and $V$ photometry. Different from our result, \citet{fah18} found a bimodal age distribution
by isochrone fitting to individual stars in Oh10, that half of the members are younger than 1\,Gyr
and the other older than 1\,Gyr. Although a WD is found in group-X ({\it Gaia} ID, 
1648892576918800128, hereafter G164), no study or observations have been carried out on it yet.

\subsection{Cluster Mass Function}\label{sec:mass}

\begin{figure}[tb!] \centering
	\includegraphics[width=\columnwidth]{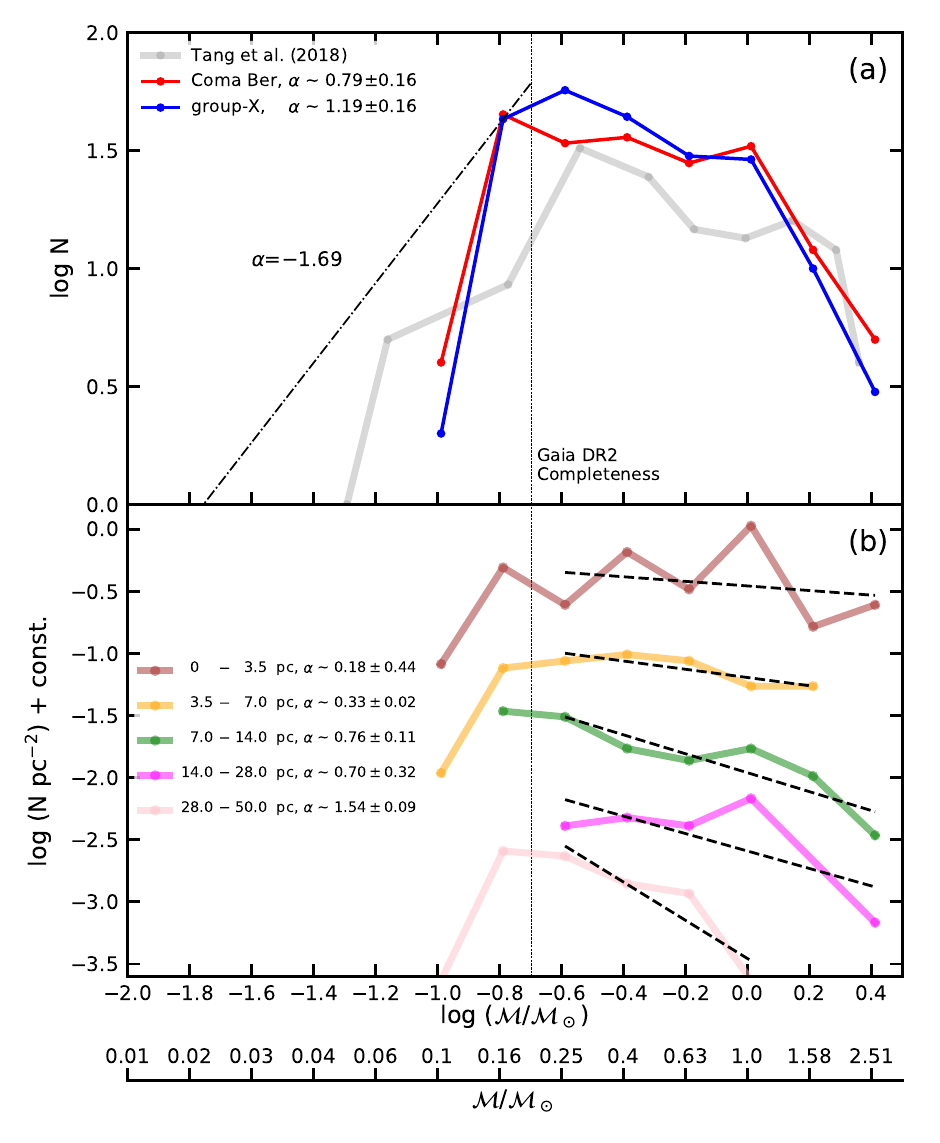}
	\caption{
	The present day mass function of Coma Ber and group-X with mass derived from PARSEC 
	isochrone with an age of 800~Myr and 400~Myr, respectively.
	(a): MFs of Coma Ber (red line) and group-X (blue line).
	    The mass function from \citet{tan18} for Coma Ber is shown in grey line.
	    The slope, $\alpha$, is fitted for MF from $0.25\,{\rm M_\sun}$ to $2.51\,{\rm M_\sun}$.
	(b): surface number density mass functions of Coma Ber in different annuli on the X--Y plane. 
	    The slope, $\alpha$, is fitted with the mass range as the black dashed lines showed.
	    The binwidth is log($M/M_\sun$)=0.2. The x-avlue of each dot represents the central 
	value of each bin. 
	}
  \label{fig:mf}
\end{figure}

By adopting an age of 800~Myr for Coma Ber, we estimate the stellar mass for each member by using 
the PARSEC isochrone, and derive the mass function (MF, Figure~\ref{fig:mf}~(a)). The slope, in a 
sense of $dN/dm \propto m^{-\alpha}$, is fitted by linear least-squares fitting. In Figure~\ref{fig:mf}~(a),
the number of low-mass members of Coma Ber increases by a factor of 5 compared to the study of 
\citet{tan18} at the mass bin of $\sim 0.16\,{\rm M_\sun}$. This to be compared with the peak around 
0.3~${\rm M_\sun}$ common seen in clusters or associations \citep{bas10}. 
The slope for present-day MF of Coma Ber ($0.25\,{\rm M_\sun}$ to $2.51\,{\rm M_\sun}$) is 
$\alpha \sim 0.79\pm0.16$, consistent with that reported by \citet{kra07} 
($\alpha \sim 0.6\pm0.3$ for $0.1-1.0\,{\rm M_\sun}$). It is flatter than that of group-X 
with $\alpha \sim 1.19\pm0.16$.

We add up the mass of member candidates in Coma Ber to obtain the cluster mass, and correct the
incompleteness below $0.2\,{\rm M_\sun}$ by adopting the slope ($\alpha\sim-1.69$, black dashed 
line in Figure~\ref{fig:mf}~(a)) from \citet{tan18} since their MF is complete down to 
$0.08\,{\rm M_\sun}$. After integrating the mass lower than $0.2\,{\rm M_\sun}$ with $\alpha\sim-1.69$, 
the mass below the completeness is $6.2~{\rm M_\sun}$. Consequently, 
the cluster mass of Coma Ber is $115\,^{+5}_{-3}\rm\,M_\sun$, with the error computed from the
uncertainty in the slope of mass function. This is
in a good agreement with previous results of $102-112\,{\rm M_\sun}$ \citep{cas06,kra07,tan18}. 
The total mass of group-X, after the same completeness correction, is
$111\,^{+16}_{-7}\rm\,M_\sun$ using the PARSEC isochrone of 400~Myr for the mass of each individual
member.

\subsection{Tidal Radii}\label{sec:radii}

The total stellar mass of Coma Ber or group-X approaches the lower limit of open cluster mass, which is around a hundred solar masses \citep{tad02,pis08}. However, either system is distributed in a large space volume 
(Figure~\ref{fig:sp_vol}), 
($80$\,pc$\times$80\,pc$\times$60\,pc) for Coma Ber and
($40$\,pc$\times50$\,pc$\times50$\,pc) for group-X. 
That is to say, the low-mass stars located in the tidal tails (especially Coma Ber) probably 
not be bound to the cluster. 

The tidal radius of a system can be computed via
\begin{equation}
r_t=(\frac{GM_C}{2(A-B)^2})^{\frac{1}{3}},
\end{equation}
\citep{pin98} where $G$ is the gravitational constant, $M_C$ is the total mass of the cluster,
and $A$ and $B$ are the Oort constants 
\citep[A$=15.3\pm0.4\rm~km~s^{-1}~kpc^{-1}, B=-11.9\pm0.4\rm~km~s^{-1}~kpc^{-1}$;][]{bov17}

With a photometric mass of $115\,^{+5}_{-3}\rm\,M_\sun$, we compute the tidal radius of Coma Ber as
$6.9\,\pm 0.1\,\rm pc$, which agrees with the values given by \cite{cas06} ($6.5$~pc) and \citet{kra07}
($6.8\pm0.3$~pc). We consider that stars within the tidal radius are bound member candidates (see
Table~\ref{tab:coma}), while stars outside belong to the tidal tails. The tidal tail at 
$X>-8300$\,pc (toward Galactic center) and $Y>0$ is leading and the other tail at $X<-8300$\,pc 
and $Y<0$ is trailing (Figure~\ref{fig:sp_vol}~(a)). 

We redetermine the center of Coma Ber as the mean position of member candidates within the tidal 
radius to be R.A.= 186.8110~deg and decl.= 25.8112~deg. These are different from those provided by 
\citet{dia14} (R.A.= 186.25~deg, decl. = 26.1~deg). With the average distance of 85.8 pc again 
obtained from member candidates within the tidal radius, the corresponding Cartesian Galactocentric 
coordinates are ($X, Y, Z$) = ($-8305.6, -5.9, +112.3$) pc. The following analysis is based on the 
revised cluster center. 

For group-X, we obtain a tidal radius of $6.8\,^{+0.3}_{-0.2}\,\rm pc$. However, 
due to its irregular morphology, determining its center is not straightforward. For simplification, 
we define its center as the mean position of all member candidates to be R.A. = 217.5175~deg, 
decl. = 55.0510~deg, and a distance of 101.2~pc, or ($X, Y, Z$) = ($-8307.2, +55.3, +109.9$) pc. 
This center lies in the gap between the two clumps and is a rough estimation. The clumpy shape of 
group-X suggests the whole system is at a final stage of disruption, since a bound and longer-lived 
structure should be roundish.  

\subsection{Dynamical Status}\label{sec:dynamical_status}

To further quantify the dynamical status of our stellar groups, we compare the photometric mass 
with the dynamical mass ($M_{dyn,tid}$) for stars within the tidal radius; i.e., these stars are
considered to be bound. Due to the irregular morphology of group-X,
we do not estimate the dynamical mass of group-X.

The dynamical mass calculated by,
\begin{equation}
    M_{dyn}\sim\frac{r_{t}\sigma_{3D}^2}{G}, 
\end{equation}
\citep{fle06} where $r_{t}$ is the 3D tidal radius, and $\sigma_{3D}$ is the 3D velocity dispersion. 
Assuming an isotropic velocity distribution within the tidal radius, 
$\sigma_{3D}^2$ is 3 times the 1D velocity dispersion $\sigma^2$ estimated from the PM. The
resulted $M_{dyn,tid}$ of Coma Ber is in the range $185\,{\rm M_\sun}$ to $375\,{\rm M_\sun}$ 
($\sigma_{\mu_\alpha \cos\delta}=0.20\,\rm~km~s^{-1}$, $\sigma_{\mu_\delta}=0.28\,\rm~km~s^{-1}$
by adopting individual distance of member candidates). Typical error of the 1D proper motion among 
member candidates is 0.1\,${\rm mas~yr^{-1}}$, corresponding to 0.04\,$\rm km~s^{-1}$, which is 
much smaller than the measured dispersion within the tidal radius. Compared to the photometric mass 
within tidal radius $\sim57\,{\rm M_\sun}$, this is a clear evidence that the cluster is
disrupting, which is an interplay between internal and external dynamical evolution. 

The major internal dynamical process, two-body relaxation, makes the high-mass stars segregate to
the cluster center while low-mass members migrate to the outskirt regions. This process manifests
as a change in the slope of the cluster MF in different annuli \citep{ves10,web15}. As can be seen
in Figure~\ref{fig:mf}~(b), the slope of radial MF becomes steeper as the radius increases,
confirming the mass segregation as discussed in Section~\ref{sec:coma}. Majority of low-mass member
candidates are located in the tidal tails, the outer extended region. Eventually, these stars will
expand and become unbound, being stripped away by the external tidal force.

\begin{figure*}[tb!]\centering
	\includegraphics[width=.8\textwidth]{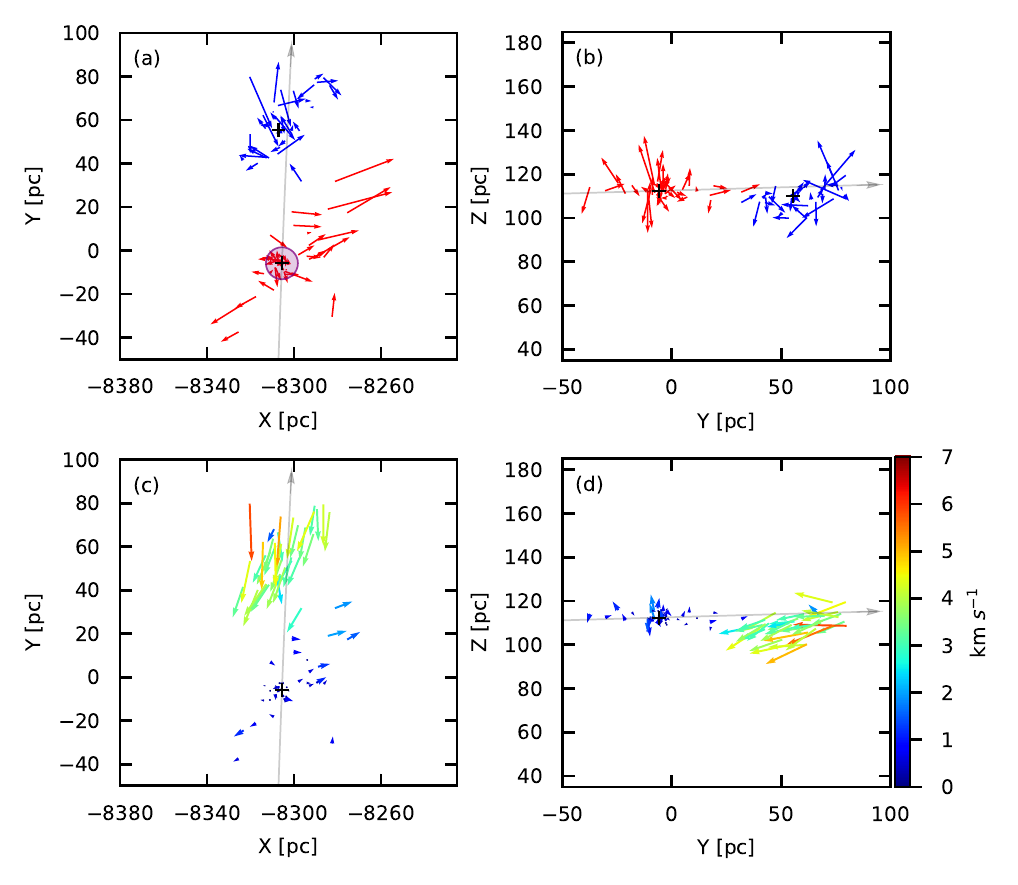}
	\caption{
	 The relative velocity vectors for member candidates in Coma Ber and group-X. 
	(a) The red vectors represent the relative velocity of member candidates of Coma Ber  
	    to its mean motion, (U, V, W)=(8.6, 226.6, 6.8)~$\rm~km~s^{-1}$ on the X--Y plane, 
	    and the blue vectors the relative velocity of member candidates of group-X to its mean 
	    motion,
	    (U, V, W)=(7.7, 223.2, 5.8)~$\rm~km~s^{-1}$;  Only stars within 1 sigma from the mean value 
	    are shown in (a) and (b). The plus sign shows the center of each group (Section~\ref{sec:radii}). 
	    Purple circle denotes the tidal radius of Coma Ber. 
	(b) the same relative velocity distribution of Coma Ber and group-X on Z--Y plane. 	
	(c) The relative velocity distributions of Coma Ber and group-X to
	    the mean motion of Coma Ber on the X--Y plane; 
	(d) the same relative velocity distribution of Coma Ber and group-X on Z--Y plane.
	The vector length and color coding are scaled to the relative velocity.
	An integrated orbit the same as the Figure~\ref{fig:sp_vol} is shown as well.
	}
	
  \label{fig:quiver}
\end{figure*}

With 3D motions (combining PM and RV), we are able to look for the tidal dissolution in Coma Ber.
We use the new cluster center (Section~\ref{sec:radii}) and mean velocity within tidal radius as
the reference frame, and present the relative 3D velocity in Figure~\ref{fig:quiver}~(a) and (b).
As can be seen, expansion is not obvious for the bound stars within tidal radius, but only
significant at tidal tail regions. Tidal-tail stars are influenced by the Galactic tides and moving 
away from the cluster center toward the outskirt, which is a clear evidence of dissolution
\citep{por01}. Similarly, member candidates of the disrupted group-X also exhibit an overall
expansion (blue arrows in Figure~\ref{fig:quiver}~(a) and (b)). 

The existence of expansion in Coma Ber may generate anisotropy in the velocity field. We compute the 
projected tangential $\sigma_T$ and radial $\sigma_R$ velocity dispersions (mean cluster motion
subtracted) within different annuli and present them in Table~\ref{tab:anisotropy}. $\sigma_T$/$\sigma_R$
= 1 implies isotropic velocity distribution. Simulations have shown that
initially isotropic star cluster will develop anisotropy as a result of two-body relaxation
\citep[$\sigma_T$/$\sigma_R<$ 1;][]{tio16} or tidal field 
\citep[$\sigma_T$/$\sigma_R>$ 1;][]{spu96,bau03,hur12}, in which radial and tangential velocity is
dominant respectively. As shown in Table~\ref{tab:anisotropy}, velocity distribution is generally
isotropic inside the tidal radius, confirming our assumption for equation (2). A weak tangential
anisotropy is found in tidal tail regions, which might be explained by the loss of unbound stars
expanding to the Galactic field \citep{lee06}.

\begin{deluxetable}{c RRR}
\tablecaption{Anisotropy of Coma Ber\label{tab:anisotropy}
             }
\tablewidth{\dimen1000}
\tablehead{
	 \colhead{Annuli}		                          & 
	 \colhead{$\sigma_{T}$}       &   \colhead{$\sigma_{R}$} &
	 \colhead{$\frac{\sigma_{T}}{\sigma_{R}}$} \\
	 \colhead{(pc)} & \multicolumn{3}{c}{(km/s)} \\
	 \colhead{(1)} & \colhead{(2)} & \colhead{(3)} & \colhead{(4)} 
	 }
\startdata
 0.0 --- 3.5   & 0.28 & 0.27 & 1.06 \\
 3.5 --- 7.0   & 0.29 & 0.29 & 0.99 \\
 7.0 --- 14.0  & 0.40 & 0.32 & 1.20 \\
 14.0 --- 28.0 & 0.45 & 0.48 & 0.95 \\
 28.0 --- 50.0 & 1.08 & 0.95 & 1.13 \\
\enddata
\tablecomments{
    $\sigma_{T}$ and $\sigma_{R}$ is the projected tangential and radial velocity dispersion 
    in Coma Ber.
    }
\end{deluxetable}

\subsection{Past Encounters and Future Flyby}

In Figure~\ref{fig:quiver}, the spatial motions of group-X relative to the center of Coma Ber is 
presented in panel (c) and (d). Stars in group-X are moving toward the Coma Ber center at a speed
of 4--6\,$\rm~km~s^{-1}$. Considering the separation of 65\,pc between group-X and the Coma Ber, 
they may flyby each other in 10--16\,Myr \citep{gav16}.
The irregular clumpy shape of group-X cannot be formed purely due to the tidal interaction with 
Coma Ber, which is not strong enough. Instead, past close encounters might be the cause. To justify 
this hypothesis, we estimate the encounter rate in the solar neighborhood. 
The encounter rate for stellar structures depends on the volume density, geometric cross-section 
of a cluster and velocity dispersion among clusters \citep{die02}. In the solar neighbourhood within
200\,pc, there are 20 stellar structures with more than 10 members, and 42 with more than 5 \citep{oh17}. 
We measure their velocity dispersion with U, V, W velocities provided by Gaia DR2 in the range of 
11--15\,$\rm km~s^{-1}$, and adopt a typical cluster radius of 10\,pc \citep{die02}. The encounter
timescale ranges from a few hundreds Myr up to 1~Gyr. It shall be an upper limit due to the 
incompleteness of \citep{oh17}. For a more extended structure, encounter rate shall increase and
encounter timescale decrease. We suspect that originally group-X might occupy a larger volume than
Coma Ber, thereby experienced more encounters and disrupted into its current morphology. A significant
amount of mass-loss must have happened to group-X.


\section{Summary}\label{sec:Summary}

Utilizing the high-precision {\it Gaia} DR\,2 astrometry, we apply a new cluster finding method,
\textsc{StarGO}, to identify stellar structures around the Coma Ber in the 5-D phase space of
($X, Y, Z$, $\mu_\alpha \cos\delta, \mu_\delta$). Additionally to the previously published member 
candidates of Coma Ber, here we identify for the first time, a leading and a trailing tidal tails 
of Coma Ber, which extend $\sim50$~pc from the cluster center, seven times longer than cluster 
tidal radius, $\sim6.9$~pc. 


\begin{enumerate}

\item The stars located in the tidal tails of Coma Ber (120) outnumber those inside the tidal
radius (77 stars). Member stars inside the tidal radius has an average distance of 85.8 pc, with
the center at R.A.= 186.8110~deg, and decl.= 25.8112~deg, corresponding to a Cartesian
Galactocentric coordinates of ($X, Y, Z$) = ($-8305.6, -5.9, +112.3$) pc.

\item Coma Ber has an age of 700--800\,Myr. It has a mass-function slope of 
$\alpha \sim 0.79\pm0.16$ for members from $0.25\,\rm{M_\sun}$ to $2.51\,{\rm M_\sun}$, and a total
cluster mass of $\sim115^{+5}_{-3}\,\rm{M_\sun}$.

\item Prominent mass segregation is seen among member candidates, but no obvious expansion is found
among stars within the tidal radius of Coma Ber. However, significant expansion is seen in the
tidal tails, with a weak anisotropy. Coma Ber is disrupting by losing stars into the Galactic
field. 

\item A neighboring group, group-X, is re-discovered with 10 times more member candidates than
previously known. Group-X is about 65 pc away from Coma Ber. With a comparable number of member
candidates as Coma Ber (197), group-X (218) is 400\,Myr younger, a steeper slope for the mass
function $\alpha \sim 1.19\pm0.16$, for stars from $0.25\,{\rm M_\sun}$ to $2.51\,{\rm M_\sun}$.

\item Group-X, with its irregular morphology, is near the end of disruption. It has the center of
R.A. = 217.5175~deg, decl. = 55.0510~deg, and a heliocentric distance of 101.2~pc, and
($X, Y, Z$) = ($-8307.2, +55.3, +109.9$)~pc.

\item Relative motion between both groups suggests a possible fly-by in the next 10--16\,Myr. 
\end{enumerate}

\acknowledgments
S.Y.T. and W.P.C. acknowledge the financial support of the grants 
MOST 106-2112-M-008-005-MY3 and MOST 105-2119-M-008-028-MY3. 
S.Y.T. also express his gratitude to the financial support of the 
summer internship at the Max Planck Institute for Astronomy in 2018.
X.Y.P. is grateful to the financial support of two grants of National Natural Science Foundation
of China, No: 11503015 and 11673032.
This study was supported by Sonderforschungsbereich SFB 881 
``The Milky Way System'' (sub-project B2, B5, and B7) of the German Research Foundation (DFG). 
Z.Y. is partly supported by the National Key Basic Research and Development Program of China 
(No.~2018YFA0404501) and NSFC grant 11673083, the Special Funding for Advanced Users through LAMOST FELLOWSHIP.
B.S. is a fellow of the International Max Planck Research School for Astronomy and Cosmic Physics at 
the University of Heidelberg (IMPRS-HD), and acknowledges the support of the Volkswagen Foundation under 
the Trilateral Partnerships grant 90411, and the PCF program BR05236322.
C.C.L. acknowledge the financial support of the grants MOST 106-2917-I-564-042.

We are grateful to Prof. Dr. Chenggang Shu and Prof. Dr. Zhengyi Shao for an in-depth discussion, 
and to Dr. Shiyin Shen for suggestion on the collaboration. We thank the referee for providing 
helpful suggestions and comments that improved the quality of the paper.
This work made use of data from the European Space Agency (ESA) mission {\it Gaia} 
(\url{https://www.cosmos.esa.int/gaia}), processed by the {\it Gaia} Data Processing and Analysis Consortium 
(DPAC, \url{https://www.cosmos.esa.int/web/gaia/dpac/consortium}). This study also made use of the SIMBAD 
database and the VizieR catalogue access tool, both operated at CDS, Strasbourg, France.

\software{Astropy~\citep{ast13,ast18}, galpy~\citep{bov15}
}

{}

\appendix{}

\section{coordinate system definitions}

The Galactocentric coordinates used in this study have a positive $x$ direction pointing from the
position of the Sun projected to the Galactic mid-plane to the Galactic center (GC) (approximately 
where $l=0$ and $b=0$); the $y$--axis points towards $l=90\degr$; the $z$--axis roughly points towards
$b=90\degr$. Other parameters are listed below:
\begin{enumerate}
\item GC coordinates (ICRS) \citep{bru04}:
    \begin{enumerate}
    \item R.A.= 17$^{\rm h}$45$^{\rm m}$37$\fs224$
    \item decl.= $-$28\degr56\arcmin10$\farcs$23 
    \end{enumerate}
\item Distance to the GC = 8.3 kpc \citep{gil09}
\item Distance of the Sun above the Galactic mid-plane = 27 pc \citep{che01}
\item Solar motion relative to the GC \citep{sch10, bov15}: \\
      (U, V, W) = (+11.10, +232.24, +7.25)~km~s$^{-1}$
\end{enumerate}

\end{document}